\documentclass[preprint,12pt]{elsarticle}

\usepackage[utf8]{inputenc}
\usepackage{graphicx}
\usepackage{amssymb}
\usepackage{placeins}
\usepackage{float}
\usepackage{lineno}
\usepackage{amsmath}
\usepackage{xcolor}
\journal{Chaos, Solitons \& Fractals}

\begin{document}

\begin{frontmatter}


\title{Mittag-Leffler functions in superstatistics}



\author{Maike A. F. dos Santos$^{1}$}

\address{$^{1}$ Centro Brasileiro de Pesquisas F\'isicas and National Institute of Science and Technology for Complex Systems, Caixa Postal 15051, CEP 91501-970, Rio de Janeiro, RJ, Brazil}

\begin{abstract}
Nowadays, there is a series of complexities in biophysics that require a suitable approach to determine the measurable quantity. In this way, the superstatistics has been an important tool to investigate dynamic aspects of particles, organisms and substances immersed in systems with non-homogeneous temperatures (or diffusivity). The superstatistics admits a general Boltzmann factor that depends on the distribution of intensive parameters $\beta=\frac{1}{D}$ (inverse-diffusivity). Each value of $\beta$ is associated with a local equilibrium in the system.  In this work, we investigate the consequences of Mittag-Leffler function on the definition of $f(\beta)$-distribution of a complex system. Thus, using the techniques belonging to the fractional calculus with non-singular kernels, we constructed a distribution to $\beta$ using the Mittag-Leffler function. This function implies distributions with power-law behaviour to high energy values in the context of  Cohen-Beck superstatistics. This work aims to present the generalised probabilities distribution in statistical mechanics under a new perspective of the  Mittag-Leffler function inspired in Atangana-Baleanu and Prabhakar forms.
\end{abstract}


\end{frontmatter}


\section{Introduction}
\label{S:1}

The statistical mechanics theory introduced by  Boltzmann and Gibbs was very successful on characterisation of macroscopic quantity by use of microscopic quantities  \cite{tolman1979principles}. One of the most important quantities to emerge from this theory is the entropy, particularly for the microcanonical ensemble. Entropy was written by Boltzmann as  $S=k_B\ln \Omega$ in which $k_B$  is the  Boltzmann constant and $\Omega$ is the number of system configurations for a fixed energy $E$. Over the years, statistical mechanics has been investigated in  contexts that transcend  the physics, such as genetic algorithm \cite{prugel1994analysis}, information theory \cite{jaynes1957information}, stochastic process in cell biology \cite{bressloff2014stochastic}.  However, the advent of experimental techniques brought a series of systems called \textit{complex systems} that required even more sophisticated theoretical formulations. In this scenario, the search for formalisms that embody more complexities in nature has attracted the attention of many scientists. Particularly, the physics of biological systems has a number of complexities. Even taking statistical approaches to address such systems, there are a number of intrinsic factors of these systems that cannot be disregarded. For example: In an equilibrium homogeneous fluid it has a temperature  $T$ and a diffusivity $D$ (in which $D\propto T$). This characteristic ceases to be valid in heterogeneous  involvements because there is fluctuation to the diffusivity $D$ (or in temperature), so such systems are out of equilibrium. They have been reported in the movement of nematodes immersed in heterogeneous environment  \cite{hapca2008anomalous}, heterogeneous random walks \cite{metzner2015superstatistical}, etc.

Thus, the theory that has been successful in addressing systems in complex environments is superstatistics  \cite{beck2003superstatistics,cohen2004superstatistics}. It provides a simple way to statistically approach the movement of a particle or organisms in complex environment. Some of the newer systems addressed through superstatistics include Brownian yet non-Gaussian diffusion \cite{chechkin2017brownian}, global warming  \cite{YALCIN20135431}, superstatistics of Fermi-Dirac and Bose-Einstein  \cite{ourabah2018fractional}, run-and-tumble particles \cite{sevilla2019stationary}, random diffusivity \cite{sposini2018random}, ultracold Buffer gas \cite{rouse2017superstatistical}, entropic forms \cite{tsallis2003constructing}, nanoscale electrochemical systems \cite{garcia2011superstatistics}, transformation groups of superstatistics \cite{hanel2011generalized}  and others.

The superstatistics was introduced by Beck and Cohen in Ref. \cite{beck2003superstatistics,abe2007superstatistics}. This theory considers the system as inhomogeneous in the intensive parameter  $\beta$, inverse of the temperature so that each different value for the intensive parameter has a particular Boltzmann factor  $e^{-\beta E}$. Therefore, superstatistics considers that the $ \beta $ value collection can be treated statistically, i.e. there is a $ f (\beta) $ distribution of the $ \beta $ intensive parameter, in which for $ t \gg T $ (long-term run) we have more general distributions that include Boltzmann factor as a particular case. Hence, being $E$ the energy associated with  microstates of a $ \beta$ inverse-temperature cell and using the above assumptions, we have
\begin{eqnarray}
\mathcal{B}=\displaystyle\int_{0}^{\infty}d\beta f(\beta) e^{-\beta E}, \label{def1}
\end{eqnarray}
which is the generalized Boltzmann factor and is very useful in approaching non-equilibrium systems. It is important to mention that non-equilibrium occurs for the entire system and that locally (in each cell) the system is in equilibrium  \cite{cohen2004superstatistics}. For a system whose fluctuations in intent values $ \beta $ collapse to a single value  $\beta_0$, i.e. $f(\beta)=\delta(\beta-\beta_0)$ we retrieve the Boltzmann-Gibbs statistic. In more complex situations the  $f(\beta)$ distribution may allow long tail which implies a number of new shapes for the generalised Boltzmann factor  \cite{han2013gamma,mathai2011pathway,sebastian2015overview}. An important point to consider is how to define the probability distribution   associated with generalised Boltzmann factor, introduced in the article \cite{beck2003superstatistics,touchette2005asymptotics} as follows
\begin{eqnarray}
 p(E) =\frac{\mathcal{B}(E)}{\mathcal{Z}}, \label{def2}
\end{eqnarray}
in which \begin{eqnarray}
 \mathcal{Z} = \sum_i\int_{0}^{\infty}d\beta f(\beta) e^{-\beta E_i},
\end{eqnarray}
is the partition function. In addition to this construction, the work \cite{touchette2005asymptotics} presents a superstatistics in which the partition function depends on $ \beta $. Such consideration implies another way of defining probability. The superstatistics theory is actually quite rich because by defining a $f(\beta)$ distribution you can build multiples types of different statistics, for example by defining  $f(\beta)$ as a Gamma distribution  (or $\chi^2$) from Eq. (\ref{def1}) the statistical mechanics proposed by Tsallis \cite{tsallis1988possible} are obtained. One of the favourable points to Tsallis statistic is that the generalised Boltzmann factor  ($e_q(-\beta E)$) by it is that within the specified limit of $ q \rightarrow 1 $ Boltzmann factor   \cite{beck2003superstatistics,tolman1979principles}. In this context, we consider a $f(\beta)$-distribution given by multiplication of a Mittag- Leffler function  \cite{haubold2011mittag} by a power-law function to define a generalisation of the Gamma distribution  (or $\chi^2$). The Mittag-Leffler (ML) functions constitute one of the most important tools that emerge in the context of fractional calculus, in this context some work has attempted to insert the Mittag-Leffler functions in the context of superstatistics \cite{mathai2012pathway,mathai2010mittag}.  
In this work, we build and analyse the superstatistics associated with two  $f(\beta)$-distributions built by ML functions. We investigate how the generalised distributions  $p(E)$ (see Eq. (2)) differ from the classical form contained in Boltzmann-Gibbs statistical mechanics.

The paper is outlined as follows: in section 2, we present the preliminaries concepts about Mittag-Leffler function \cite{haubold2011mittag} in context of fractional calculus. In section 3 we introduce the density-diffusivity that consists of the construction of distribution using Mittag-Leffler functions. In the following, we present a series of behaviour to exemplify the different behaviours to generalised probability distributions for two particular cases of Mittag-Leffler density-diffusivity. Finally, in section 4, we present the conclusions that include a discussion of possible scenarios that results can be applied.

\section{Preliminary concepts about Mittag-Leffler function}
\label{S:1}

The Mittag-Leffler function defined appears in a context of fractional calculus. The fractional calculus is the science fields that investigate the fractional derivative operator. There are several other definitions of fractional derivatives, and these satisfy several mathematical properties that are detailed in the reference \cite{rudolf2000applications}. The Fractional derivatives, such as Riemann-Liouville, Letnikov, Riesz, and others, are constituted by convolution integrals with power-law kernels. These derivatives applied in differential equations generate a series of special functions \cite{samko1993fractional}, the Mittag-- Leffler and Fox functions when applied in contexts associated with particles diffusion. In this context, the Mittag-Leffler function defined as follow 
\begin{eqnarray}
E_{\alpha} \left(  z \right)= \sum_{k=0}^{\infty} \frac{z^{k}}{\Gamma[\alpha k + 1 ]} ,
\label{mitag1}
\end{eqnarray}
emerges as a natural solution of equation like $\frac{d^{\alpha}}{dz^{\alpha}}y(z)=y(z)$ ($y(0)=0$ and $z \in \mathbb{R}^+$). In particular, if $\alpha \rightarrow 1$ we recover the exponential function. Still in the context of the fractional calculus there is a class of derivative operator that are non-singular and are defined as a convolution under a Mittag-Leffler function. So, the Mittag-Leffler function has been a fundamental function in fractional calculus \cite{sene2019mittag,kilbas2004generalized,gorenflo2014mittag}. In fact, the fractional derivatives describe a series of problems in complex systems \cite{dos2019analytic,oliveira2019anomalous,dos2019continuous}.

Now we consider the tempered fractional derivative with three parameters Mittag-Leffler kernel \cite{fernandez2019series,garra2014hilfer}
 in Caputo sense is defined as follows
\begin{eqnarray}
 \mathcal{D}_{\alpha, \sigma, t
}^{\delta, \nu,a} f(t)= \int_0^{t}  e^{-a (t-t')} \wp_{\alpha,\sigma}^{\delta,\nu}(t-t') \frac{d\ }{dt'} f(t') dt',
\qquad t \in \mathbb{R}, 
\label{Prabhakar}
\end{eqnarray}
with the Prabhakar kernel $\wp_{\alpha,\sigma}^{\delta,\nu}$ defined by
\begin{eqnarray}
\wp_{\alpha,\sigma}^{\delta,\nu}[t]=t^{\sigma-1} E^{\delta}_{\alpha, \sigma}\left(- \nu t^{\alpha} \right),
\label{prabhakarK}
\end{eqnarray}
in which $E_{\alpha,\sigma}^{\delta}(z)$ is the generalised Mittag - Leffler function for three parameters \cite{prabhakar1971singular}, given by 
\begin{eqnarray}
E^{\delta}_{\alpha, \sigma} \left(  z \right)= \sum_{k=0}^{\infty} \frac{(\delta)_k}{\Gamma[\alpha k + \sigma ]} \frac{z^{k}}{k!},
\label{mitag1}
\end{eqnarray}
 $(\delta)_k = \Gamma[\delta + k ] / \Gamma[\delta] $ is the Pochhammer symbol, com $\mathcal{R}\{\sigma \}>0$ e $\sigma, \alpha, \delta, z \in  \mathbb{C}$. The function (\ref{mitag1}) recoveries the two--parameters Mittag-Leffler function  \cite{podlubny1998fractional} to $\delta=1$, i.e. $E_{\alpha,\sigma}^1(z)$. The     Mittag-Leffler function (\ref{mitag1}) is reduced from one parameter to $\sigma=1$ and $\delta=1$. Finally, the function  (\ref{mitag1}) assumes the exponential form when $\alpha=\sigma=\delta=1$, i.e. $E_{1,1}^1(z)=e^{z}$.
Among the advantages of using the Prabhakar derivative (Eq. (\ref{Prabhakar})), there is the fact that the Caputo derivative \footnote{Or Riemann--Liouville, depending on how the Prabhakar derivative is defined.} is a particular case of the Prabhakar derivative. The Laplace transform of Eq. (\ref{prabhakarK}) is summarised as follows
\begin{eqnarray}
\mathcal{L} \left\{  t^{\sigma -1} E^{\delta}_{\alpha, \sigma}\left(- \nu t^{\alpha} \right)  \right\} = \frac{s^{\alpha\delta - \sigma}}{(s^{\alpha} + \nu )^{\delta}} \qquad \mathcal{R}\{ s \} < |\nu|^{\frac{1}{\alpha}},
\label{inversa1}
\end{eqnarray}
in which $\delta, \alpha, \sigma \in \mathbb{C}$ and $\mathcal{R}\{ \alpha \}, \mathcal{R} \{\sigma \} > 0$. The Eq. (\ref{inversa1}) satisfies a series of mathematical proprieties that were detailed in Ref. \cite{garra2014hilfer}.
Note that to $\delta=0$, $a=0$ and $\sigma=1$, the Eq. (\ref{Prabhakar}) retrieves the fractional derivative form of Caputo.  Thereby, the fractional Prabhakar derivative has been an efficient tool in physical models to approach the transition among anomalous diffusion \cite{sandev2018models} and stochastic resetting problem \cite{dos2019fractional}, etc. In this decade was defined a fractional derivative by Atangana and Baleanu that has been very applied in general science, the Atangana-Baleanu (AB) fractional derivative reported in \cite{atangana2016new} is defined as a convolution integration of a Mittag-Leffler function with one parameter in convolution term and an arbitrary function, i.e. $\frac{d\ }{dt}\int_a^t dt' \mathcal{K}_{AB}(t')f(t-t')$. The AB fractional derivative recovers the ordinary derivative of the first order when  $\alpha\rightarrow 1$. The AB fractional operator has revealed a class of interesting behaviours in amount quantity of problems in physics. Newer investigations include themes as chaos and statistic \cite{atangana2018fractional}, reaction-diffusion systems \cite{saad2018analysis}, time-fractional variable-order telegraph equation \cite{gomez2019time}, Lévy Flights \cite{dos2019mittag}, viscoelastic response \cite{hristov2019linear}, etc. In fact, in this scenario, the Mittag-Leffler kernels have been a mathematical tool that has been more understood day by day \cite{sene2019analysis,fernandez2019series}. Thereby, this work extends the applicability of this function to a new field in physics of complex systems, the Cohen-Beck superstatistics.

\section{Constructing a  $f(\beta)$-distribution with Mittag-Leffler function}

In this section, we build a distribution of intensive $\beta$ parameters using the Mittag-Leffler function. Furthermore it is important to emphasise that we consider the $\beta $ parameter in our development, and the diffusivity is given by the inverse relationship with the  $\beta$ parameter, i.e. $\beta=D^{-1}$, in which $D$ is the diffusivity of a cell with  $T$ temperature \cite{cohen2004superstatistics}. Now, to continue our proposal we introduce some conditions required by the superstatistics   \cite{beck2003superstatistics}, formalism, which are as follows:
\begin{enumerate}
    \item The $ f (\beta) $ distribution of the  intesive parameters $ \beta $ must be normalised, i.e. $ \int_0 ^{\infty} f (\beta) d \beta = 1 $. This guarantees a statistical treatment for theory and has relevance from the point of view of physics. \\
    \item  The probability distribution $ p (E) $ must be normalized, i.e. the integral $ \int_0 ^ {\infty} dE p (E) $ must exist. In the most general case, the integral $$\int_0^{\infty}dE p(E)g(E)$$  must exist in which $ g (E) $ is the density of states.
\end{enumerate} 
In addition to these considerations, it is important to emphasise again that for $f(\beta) = \delta (\beta- \beta_0) $ we regain Boltzmann's weight. Furthermore, this proposal may justify the foundations of non-extensive statistical mechanics proposed by Tsallis \cite{beck2001dynamical}. Inspired by the $ \chi^2$-Gamma distribution that implies the generalised Boltzmann factor by Tsallis \cite{tsallis1988possible}, let's assume the general mathematical structure for intensive parameter distribution as follows
\begin{eqnarray}
f(\beta) \propto e^{-b\beta}\beta^{\sigma-1} E_{\alpha,\sigma}^{\delta}[-a \beta^{\alpha}],
\end{eqnarray}
in which this type of structure appears in a number of contexts in physics to address tempered systems \cite{dos2019fractional}. We can use a series of well-known fractional calculation tools to enter the normalisation constant of the above equation and determine the average value $ \langle \beta \rangle $. The $ f (\beta) $ function Laplace transform is defined by $ \int_0 ^ {\infty} e^{- s \beta} f (\beta) d \beta $ and results in
\begin{eqnarray}
\mathcal{L} \left\{ f(\beta) \right\} =  \frac{(b+s)^{\alpha \delta - \sigma}}{(a+(b+s)^{\alpha})^{\delta}},
\end{eqnarray}
we can define a normalised distribution as follow 
\begin{eqnarray}
\mathcal{L}\{f(\beta) \}&=& c\mathcal{L} \left\{  e^{-b\beta}\beta^{\sigma -1} E_{\alpha, \sigma}^{\delta}[-a \beta^{\alpha}]\right\},\label{eqm}
\end{eqnarray}
the average normalise the  quantity $ f(\beta) $ is given by 
\begin{eqnarray}
\int_0^{\infty} f(\beta) d \beta = \lim_ {s \rightarrow 0} \int_0 ^ {\infty} e^ {- s \beta} f (\beta) d \beta,
\end{eqnarray}
 which is a limit on the Laplace transform. Thus, using Eq. (\ref{eqm}) and the definition of mean, we have the following normalisation constant $ c = (a + b ^ {\alpha}) ^ {\delta} (b ^ {\alpha \delta - \beta}) ^{-1} $ which implies the following distribution
\begin{eqnarray}
f(\beta) = \frac{(a+b^{\alpha})^{\delta}}{b^{\alpha \delta -\sigma}}e^{-b\beta}\beta^{\sigma-1} E_{\alpha,\sigma}^{\delta}[-a \beta^{\alpha}],
\label{defini}
\end{eqnarray}
in which $\beta=D^{-1}$.

We can now set the average value of $ \langle \beta \rangle $ to ensure that it is a finite quantity. Using the same technique as the Laplace transform, we can write the quantity $ \mathcal{L} \{t f (t) \} $ so
\begin{eqnarray}
\mathcal{L}\{\beta f(\beta) \}&=&\mathcal{L} \left\{ \beta \frac{(a+b^{\alpha})^{\delta}}{b^{\alpha \delta -\beta}} \beta^{\alpha \delta -1} E_{\alpha,\alpha \delta}^{\delta}[-a \beta^{\alpha}] \right\} \nonumber \\ &=& -\frac{d\ }{ds} \frac{(s+b)^{\alpha \delta - \beta}}{(a+(s+b)^{\alpha})^{\delta}} \frac{(a+b^{\alpha})^{\delta}}{b^{\alpha \delta -\beta}}  \nonumber  \\
&=& \frac{(a+b^{\alpha})^{\delta}}{b^{\alpha \delta -\beta}}  \frac{\delta \alpha (s+b)^{\alpha-1}}{(a+(s+b)^{\alpha})^{\delta+1}} (s+b)^{\alpha \delta - \beta} \nonumber \\ &-& \frac{(a+b^{\alpha})^{\delta}}{b^{\alpha \delta -\beta}}  \frac{\delta \alpha -\beta }{(a+(s+b)^{\alpha})^{\delta}} (s+b)^{\alpha \delta - \beta-1}.
\end{eqnarray}
Then, we can determine exactly the average value of the $ \beta $ parameter is given by $\langle \beta \rangle = \lim_{s \rightarrow 0} \int_0^{\infty}e^{-s \beta} \beta f(\beta) d\beta $, we obtain
\begin{eqnarray}
\langle \beta \rangle &=& \lim_{s \rightarrow 0}\mathcal{L}\{\beta f(\beta) \} \nonumber \\ &=&  \frac{(a+b^{\alpha})^{\delta}}{b^{\alpha \delta -\beta}}  \frac{\delta \alpha b^{\alpha-1}}{(a+b^{\alpha})^{\delta+1}} b^{\alpha \delta - \beta} \nonumber \\  &-& \frac{(a+b^{\alpha})^{\delta}}{b^{\alpha \delta -\beta}}  \frac{\delta \alpha -\beta }{(a+b^{\alpha})^{\delta}} b^{\alpha \delta - \beta-1} \nonumber \\ &=&  \frac{\delta \alpha b^{\alpha-1}}{a+b^{\alpha}} - (\alpha \delta - \beta) b^{-1},\label{media1}
\end{eqnarray}
in which $ \langle \beta \rangle> 0 $. Here, we can divide the results into two parts. The first of these refers to the distribution associated with a $ f (\beta) $ distribution with kernel characteristics defined by Atangana and Baleanu \cite{atangana2016new}. The second part concerns the particular parameter $\sigma = \alpha \delta$. These two cases satisfy the required conditions for $ f (\beta) $ to be associated with a superstatistics.

\section{First case: One parameter Mittag-Leffler function for $f(\beta)$ distribution}

\begin{figure}[h!]
\centering
\includegraphics[width=0.65\textwidth]{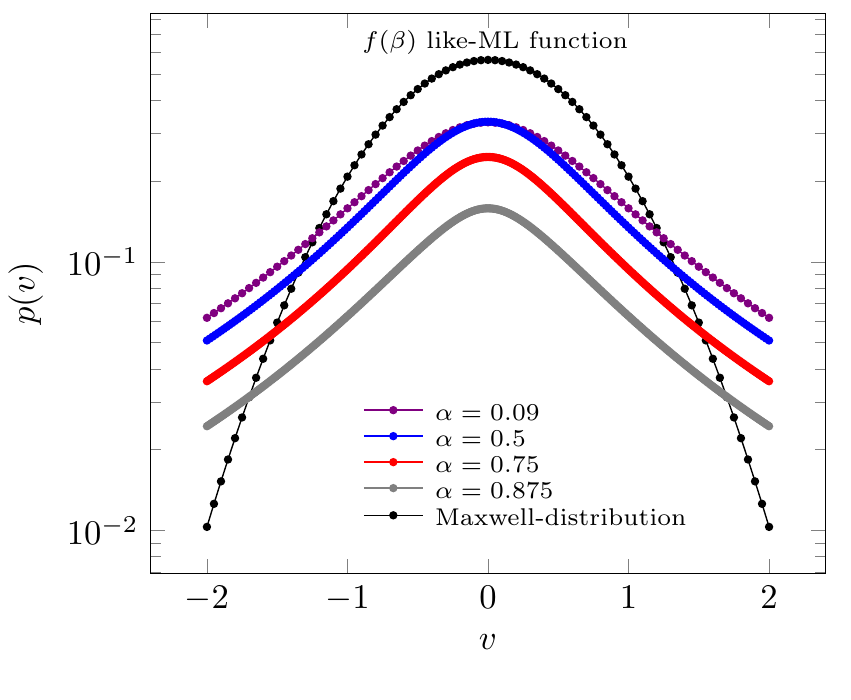} 
\caption{This figure show a series of curves that represent some particular situations of Eqs. (\ref{likemax}) and (\ref{Maxwelldistribution}) with follow parameters: $\beta_0=1$, $m=2$, $b^{-1} = \beta_0(1-\alpha)^{-1}$ and different $\alpha$-index.}
\label{fig1}
\end{figure}

The first case consists of a detailed analyse of a superstatistics that is based on the kernel proposed by Atangana and Baleanu in the article \cite{atangana2016new}, so we have the following conditions:
 \begin{eqnarray}
 \delta&=&\sigma=1 \\
 a &=& \frac{\alpha}{1-\alpha},
 \end{eqnarray} 
in Eq. (\ref{defini}) we obtain the following mathematical expression
 \begin{eqnarray}
f(\beta) = \left(\frac{\alpha (1-\alpha)^{-1} +b^{\alpha}}{b^{\alpha  -1}}\right)e^{-b\beta}\beta^{\beta-1} E_{\alpha}\left[-\frac{\alpha}{1-\alpha} \beta^{\alpha}\right]. \label{firstcaseF}
\end{eqnarray}
Using Eq. (\ref{media1}) and considering $ \langle \beta \rangle> 0 $ we have
\begin{eqnarray}
 \frac{ \alpha b^{\alpha}}{\alpha (1-\alpha)^{-1}+b^{\alpha}}  > (\alpha - 1),
\end{eqnarray}
so 
$1- \frac{\alpha (1-\alpha)^{-1}}{\alpha (1-\alpha)^{-1}+b^{\alpha}}  > (1 - \alpha^{-1})$ that implies 
\begin{eqnarray}
\frac{1}{\alpha}  > \frac{ \alpha (1-\alpha)^{-1}}{\alpha (1-\alpha)^{-1}+b^{\alpha}}.\label{desigualdade1} 
\end{eqnarray}
Therefore, 
$\alpha (1-\alpha)^{-1} + b^{\alpha} > \alpha^2 (1-\alpha)^{-1}$ is the condition that must satisfied. Therefore we consider  $1>\alpha >0$. Continuing our approach related to the Atangana-Baleanu kernel, we have $ a = \frac{\alpha}{1- \alpha} $ which $ 1> \alpha> 0 $. Then, replacing (\ref{firstcaseF}) in (\ref{def2}) we have
\begin{eqnarray}
p(E) &=& \frac{1}{ \mathcal{Z}_\alpha}  \frac{(b+E)^{\alpha - 1}}{\left(\frac{\alpha}{1-\alpha}+(E+b)^{\alpha} \right)}, \end{eqnarray}
in which $\mathcal{Z}_{\alpha}^{-1}=\mathcal{Z}^{-1}\frac{(\frac{\alpha}{1-\alpha}+b^{\alpha})}{b^{\alpha  - 1}}$, so that
\begin{eqnarray}
p(E) &=& \frac{\displaystyle (\alpha+(1-\alpha)b^{\alpha})}{ \mathcal{Z}}  \frac{(1+b^{-1}E)^{\alpha - 1}}{\displaystyle \alpha+(1-\alpha)b^{\alpha}(1+b^{-1}E)^{\alpha} }, \end{eqnarray}
to $b^{-1} = \beta_0(1-\alpha)^{-1}$ and knowing that $a=\frac{\alpha}{1-\alpha}$ a
%(\alpha-1)^{-1} &=& - x  \qquad 1>\alpha>0,
 the relation (\ref{desigualdade1}) is satisfied. Thus, we have
\begin{eqnarray}
p(E) &=& \frac{(b^{-\alpha}\frac{\alpha}{1-\alpha}+1)}{ \mathcal{Z}}  \frac{(1+(1-\alpha)^{-1} \beta_0  E)^{\alpha-1}}{ \displaystyle \frac{\alpha b^{-\alpha}}{1-\alpha}+(1+(1-\alpha)^{-1} \beta_0 E)^{\alpha} }, \label{result1E}  %\nonumber \\ 
%&=& \frac{((x\beta_0 )^{(1-\frac{1}{x})} x (1-\frac{1}{x}) +1)}{ \mathcal{Z}} \times  \nonumber \\ &\times & \frac{(1+x \beta_0  E)^{-\frac{1}{x}}}{(x \beta_0 )^{(1-\frac{1}{x})}(1-\frac{1}{x}) x  +(1+ x\beta_0 E)^{(1-\frac{1}{x})}}
\end{eqnarray}
in which $ \mathcal{Z} $ is the partition function that normalises the probability.
An asymptotic analysis of Eq. (\ref{result1E}) can be done with this result, the first limit occurs for high energies
\begin{equation}
    \lim_{E\rightarrow \infty} p(E) \sim \frac{1}{E}, 
\end{equation}
the second limit for low power, i.e. $ E \rightarrow 0 $ we obtain $ p (E) \sim \mathcal{Z} ^ {- 1} $. We can write the probability distribution $ p (E) $ as a function that depends on kinetic energy and potential energy. So, in phase space (changing the notation $ E $ to $ \mathcal{H} $) we have to
\begin{eqnarray}
p(\vec{v},\vec{r}) &=& \frac{(b^{-\alpha}\alpha j+1)}{ \mathcal{Z}}  \frac{\displaystyle \left(1+j \beta_0  \displaystyle \frac{m\vec{v}^2}{2}+j \beta_0 V(\vec{r})\right)^{\alpha-1}}{\displaystyle \frac{\alpha b^{-\alpha}}{1-\alpha}+\left(1+ j\beta_0 \left\{ \displaystyle \frac{m\vec{v}^2}{2}+V(\vec{r}) \right\} \right)^{\alpha}}, \label{eq20}
\end{eqnarray}
in which $j=(1-\alpha)^{-1}$, $b^{-1} = \beta_0(1-\alpha)^{-1}$, $\int\int d\vec{v}d\vec{r}p=1$ and $\mathcal {H} (\vec{v}, \vec{r}) = m \vec{v}^2 2^{- 1} + V(\vec{r}) $. Here we can assume that the system is not about the action of a potential, i.e. $ V (\vec{r}) = 0 $. For simplicity, we can also consider that the generalised one-dimensional canonical \textit{ensemble} (like a box with $ L $ measurement) so Eq. (\ref{eq20}) is reduced as follows
\begin{eqnarray}
p(\vec{v}) &=& \frac{(b^{-\alpha}\alpha j+1)}{ \mathcal{Z}}  \frac{\left(1+j \beta_0  \displaystyle \frac{m\vec{v}^2}{2}\right)^{\alpha-1}}{\displaystyle \frac{\alpha b^{-\alpha}}{1-\alpha}+\left(1+ j\beta_0 \displaystyle \frac{m\vec{v}^2}{2}\right)^{\alpha}},\label{likemax}
\end{eqnarray}
in which $j=(1-\alpha)^{-1}$ and $b^{-1} = \beta_0(1-\alpha)^{-1}$. This result describes the velocity distribution of particles immersed in a medium with the distribution of $ \beta $ parameter governed by a Mittag-Leffler function with the same parameter choice made in the Atangana-Baleanu kernel, i.e. $ a = \frac{ \alpha}{1- \alpha} $. Considering a uni-dimensional box a parallel can be made in this text through Maxwell's distribution. Maxwell's distribution is typically a Gaussian distribution on velocity, written as
\begin{eqnarray}
p(v)= \sqrt{\frac{\beta_0 m }{2\pi}}e^{-\frac{\beta_0mv^2}{2}}. \label{Maxwelldistribution}
\end{eqnarray}
The Eq. (\ref{likemax}) does not converge to Maxwell's equation for any value of $ \alpha $, yet has a very rich behaviour class as shown in Fig. (\ref{fig1}). The type of non-Gaussian behaviour in velocity distribution is important to approach turbulent systems \cite{beck2009recent}. In the next section we observe that there is a suitable choice of parameters in the $ f (\beta) $ defined function that recovers the Boltzmann factor, and as a consequence the Boltzmann-Gibbs statistical mechanics.

\section{Second case: Like-Prabhakar kernel for  $f(\beta)$-distribution  ($\alpha \delta = \sigma$)}

\begin{figure}[H]
\centering
\includegraphics[width=0.65\textwidth]{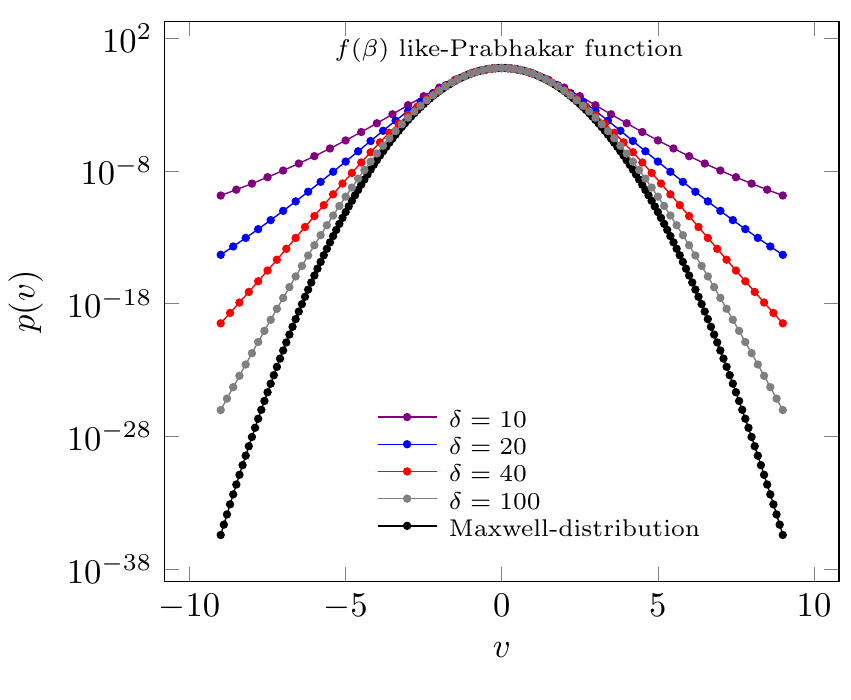} 
\caption{This figure show a series of curves that represent some particular situations of Eq. (\ref{prob3}) with $E=mv^22^{-1}$ to follow parameters: $\beta_0=1$, $m=2$, $b=0$, $\alpha=1$ and different $\delta$-index. The black-curve represent the Maxwell distribution (Eq. (\ref{Maxwelldistribution})) with $m=2$ and $\beta_0=1$.}
\label{fig2}
\end{figure}

\begin{figure}[h!]
\centering
\includegraphics[width=0.65\textwidth]{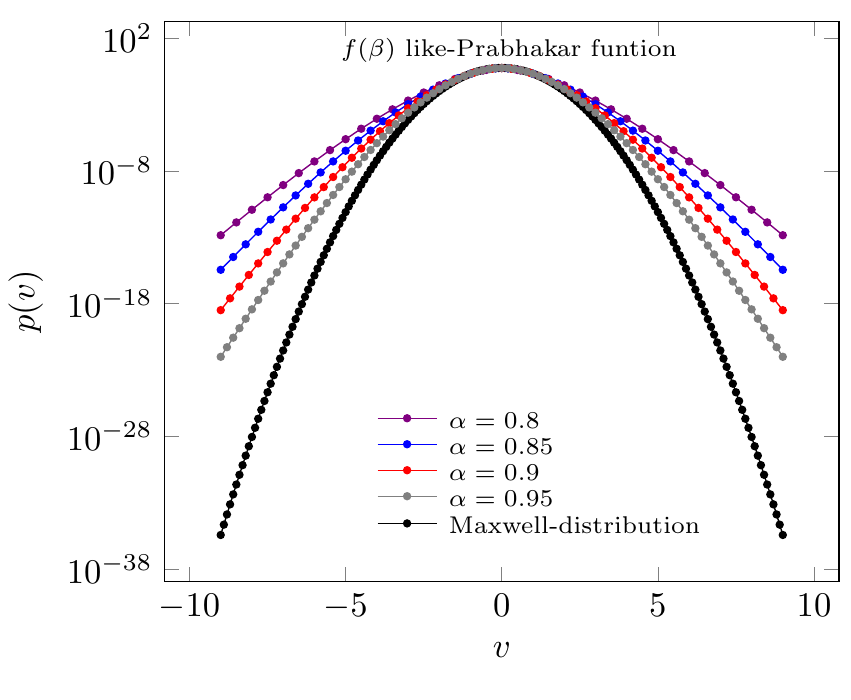} 
\caption{This figure show a series of curves that represent some particular situations of Eq. (\ref{prob3}) with $E=mv^22^{-1}$ to follow parameters: $\beta_0=1$, $m=2$, $b=0$, $\delta=10^2$ and different $\alpha$-index. The black-curve represent the Maxwell distribution (Eq. (\ref{Maxwelldistribution})) with $m=2$ and $\beta_0=1$.}
\label{fig3}
\end{figure}

\begin{figure}[H]
\centering
\includegraphics[width=0.65\textwidth]{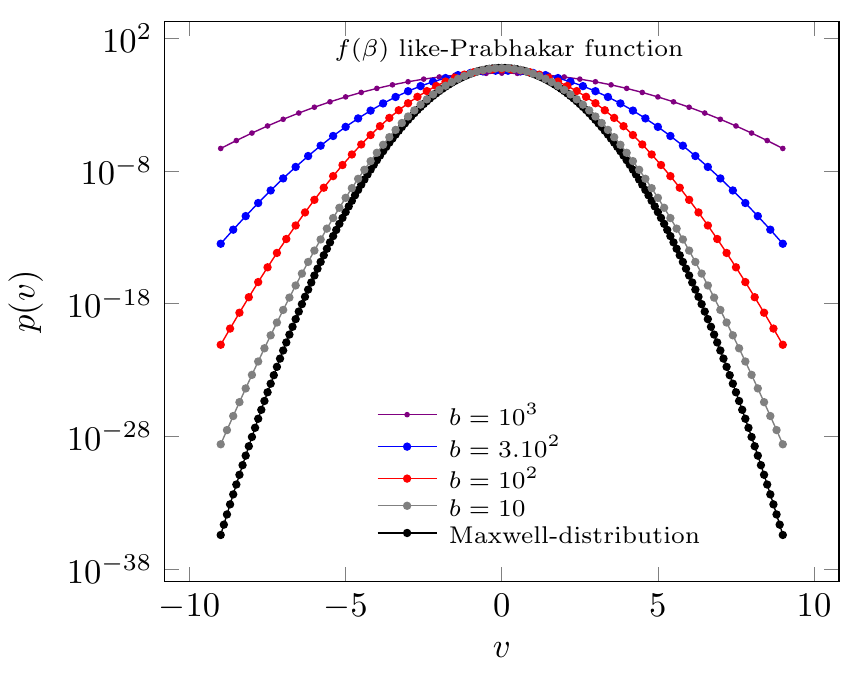} 
\caption{This figure show a series of curves that represent some particular situations of Eq. (\ref{prob3}) with $E=mv^22^{-1}$ to follow parameters: $\beta_0=1$, $m=2$, $\alpha=1$, $\delta=10^2$ and different $b$ parameter. The black-curve represent the Maxwell distribution (Eq. (\ref{Maxwelldistribution})) with $m=2$ and $\beta_0=1$.}
\label{fig4}
\end{figure}

In this section, we analyse a superstatistics associated with the Mittag-Leffler function under  $\sigma= \alpha \delta $ constraint in Eq. (\ref{defini}) which implies
\begin{eqnarray}
f(\beta) = (a+b^{\alpha})^{\delta}e^{-b\beta}\beta^{\alpha\delta-1} E_{\alpha,\alpha \delta}^{\delta}[-a \beta^{\alpha}]. \label{prab2}
\end{eqnarray}
Using the Eq. (\ref{media1}) and $\langle \beta \rangle >0$ to the Eq. (\ref{prab2}) we has  $
  \delta \alpha b^{\alpha-1}(a+b^{\alpha}) >0$
 that imply in real-positive values to $a$ and $b$. Thereby, using the Eq. (\ref{prab2}) in definitions  (\ref{def1}) and (\ref{def2}) we obtains
\begin{eqnarray}
p(E)&=&\frac{1}{\mathcal{Z}}\frac{(1+a^{-1}b^{\alpha})^{\delta}}{(1+a^{-1}(E+b)^{\alpha})^{\delta}}, \label{prab21}
\end{eqnarray} 
in which $ \mathcal{Z} $ is the partition function that normalises the probability.
Just as in the previous section we can determine the asymptotic limit of this result, the first limit occurs for high energies.
\begin{equation}
    \lim_{E\rightarrow \infty} p(E) \sim \frac{1}{E^{\alpha \delta}}. \label{assintoticoalpha}
\end{equation}
So that $ E \gg b $, for the case that the product $ \delta \alpha = 1 $ the decay is analogous to that found by the density-diffusivity presented in the previous section. Let's rewrite Eq. (\ref{prab21}) in the $ (v, r) $ space by defining the $ E \rightarrow \mathcal{H} (\vec{v}, \vec{r}) $ energy, so
\begin{eqnarray}
p(\vec{v},\vec{r})&=&\frac{\left(1+a^{-1}(\mathcal{H}(\vec{v},\vec{r})+b)^{\alpha}\right)^{-\delta}}{\mathcal{Z}_{\alpha,\delta}},\label{prob3}
\end{eqnarray} 
in which we rewrote the normalization constant $ \mathcal{Z} $ as $ \mathcal{Z} _ {\alpha, \delta} = \mathcal{Z} (1 + a ^ {- 1} b ^ {\alpha} ) ^ {- \delta} $. Considering a gas confined in a one-dimensional system, as considered in the previous section we have $ \mathcal {H} = mv^ 2 2 ^{- 1} $. Furthermore we know that for the case which the environment is not heterogeneous with respect to the intensive parameter $ \beta $ we obtain the Maxwell-distribution to velocities, for the case that there is a complex environment described by the distribution (\ref{prob3}) we obtain a series of results that for particular situations (limits on the parameters) recover the Maxwell-distribution. The Fig. (\ref{fig2}) illustrates the velocity distribution for a free particle gas subject to the probability distribution (\ref{prob3}) in the case that $ b = 0 $, $ a = \delta \beta_0 ^ {- \alpha} $, $ \alpha = 1 $ and $ \mathcal{Z} $ is the normalisation constant, in this figure we show that for low values of $ \delta $ the distribution has a long syrup which according to the relation (\ref{assintoticoalpha}) is given by $ p (v \rightarrow \infty) \sim v ^ {- 2 \delta} $. For $ \delta \rightarrow \infty $ we obtain the Maxwell distribution as the boundary case, that is, we retrieve the Boltzmann statistic. This is only because we set the value of $ \alpha = 1 $, knowing that we can now consider a high value of $ \delta $, i.e. $ \delta \gg 1 $ fixed, and changing the values of $ \alpha $, this new situation is represented in Fig. (\ref{fig3}), in which case it is clear that for $ \alpha <1 $ the system has a long tailed behaviour and only recovers the Maxwell distribution to $ \alpha \rightarrow 1 $. The third and last situation represented by Fig. (\ref{fig4}) considers the value of the parameters $ \alpha $ and $ \delta $ to be fixed and changes the amount $ b $ which for the other analyses was considered null. Thus Fig. (\ref{fig4}) shows a series of new long tailed  distributions, which correspond to after cases with  energy  plus a constant.

All of these features discussed and graphically demonstrated for the velocity distribution associated with Mittag-Leffler distribution type  to $f(\beta)$ distribution. Now, let's simply consider $ b = 0 $ to analyse a special case in Eq.(\ref{prab21}). Also let's consider that $ a = \delta \beta_0 ^ {- \alpha} $, so
\begin{eqnarray}
p(E)&=&\frac{(1+\frac{\beta_0^{\alpha}}{\delta}E^{\alpha})^{-\delta}}{\mathcal{Z}},\label{eq32}
\end{eqnarray} 
in which $\sigma= \alpha\delta$. To $\delta \rightarrow \infty$ we obtain 
\begin{eqnarray}
\lim_{\delta \rightarrow \infty}p(E)&=&\frac{e^{-(\beta_0 E)^{\alpha}}}{\mathcal{Z}},
\end{eqnarray} 
so that to $\alpha \rightarrow 1$ we recover the usual Boltzmann factor 
\begin{eqnarray}
\lim_{\substack{\delta \to \infty \\ \alpha \to 1}} \left(1+\frac{\beta_0^{\alpha}}{\delta}E^{\alpha}\right)^{-\delta}\rightarrow e^{-\beta_0 E}, \label{limitB}
\end{eqnarray}
in which the effective diffusivity and effective temperature obeys the Einstein- Smoluchowski relation
\begin{eqnarray}
\beta_0 = \frac{1}{D_{eff}} = \frac{1}{k_B T_{eff}},
\end{eqnarray}
for the Boltzmann factor obtained at the expression limit (\ref{limitB}).
The result expressed by Eq. (\ref{eq32}) is of great importance because retrieve the Boltzmann-Gibbs statistic. In addition to this result it is important to mention if we assume $ \delta = \frac{1}{q-1} $ and $ \alpha = 1 $ in Eq. (\ref{eq32}) we obtain the distribution proposed by Tsallis in the \cite{tsallis1988possible}. In other words, the result contained in Eq. (\ref{eq32}) corresponds to a statistical mechanics that reclaims the Boltzmann-Gibbs \cite{tolman1979principles} and Tsallis \cite{tsallis1988possible} statistics as special cases.

\section{Conclusion}

In this work, we proposed a Mittag-Leffler function to obtain a statistic-mechanics type to irregular (non-homogeneous) environments. Moreover, we considered the Mittag-Leffler function to build a distribution $f(\beta)$ of intensive parameters. Moreover, we investigate the consequences in the probability for two particular cases, the first one was inspired in AB kernel \cite{atangana2016new} and the second one involve a Mittag-Leffler function with three parameters (to $ \sigma = \alpha \delta$ in Eq. (\ref{inversa1})). The results found here are accurate and generate relevant consequences in complex environment.

First, we addressed the problem of ML function to $f(\beta)$-distribution for a simple case of the Mittag-Leffler function with one parameter in the sense of AB kernel. Thereby, we obtain the probability distribution and we have shown that associated with this case to high energy values the probability distribution has tailed $ p (E) \propto E^{- 1} $ that does not depend on $ \alpha $ index. Moreover, we use this result to investigate a gas of particles in a complex environment, we present a series of generalised distribution with long tail.

In the sequence, we investigated a superstatistics associated with $f(\beta)$-distribution associated to Mittag-Leffler function with three parameters. In this case, we considered a particular restriction under parameters choice $\sigma=\alpha \delta $ in three-parameters Mittag Leffler function. We obtain a probability that depends on two parameters, the particular limit to $ E \rightarrow \infty $ we obtain a long-tailed as $ p (E) \propto  E ^ {- \alpha \delta} $. We have presented the analytical expression for gas of particles that are evenly distributed in a complex environment associated with $f(\beta)$-distribution. We obtain the limit that recover the Boltzmann statistical mechanics, i.e. there are proper choices and limits of parameters that recover the Boltzmann factor.

Our research has revealed that the superstatistics brings new results associated with a $f(\beta)$-distribution  built with combination of Mittag-Leffler function and a power-law function. In addition, we show that the tools investigated can be useful in the approach of Maxwell-gas in complex environment (or disordered). Therefore, the results contribute to a new point of view in the investigation of a new statistical-mechanics of Mittag-Leffler function to $f(\beta)$-distribution.

\section*{Acknowledgements} 
M. A. F. dos Santos acknowledges the support of the Brazilian agency CAPES/INCT.



%% References


\bibliographystyle{model1-num-names}

\begin{thebibliography}{10}

\bibitem{abe2007superstatistics}
Sumiyoshi Abe, Christian Beck, and Ezechiel~GD Cohen.
\newblock Superstatistics, thermodynamics, and fluctuations.
\newblock {\em Physical Review E}, 76(3):031102, 2007.

\bibitem{atangana2016new}
Abdon Atangana and Dumitru Baleanu.
\newblock New fractional derivatives with nonlocal and non-singular kernel:
  theory and application to heat transfer model.
\newblock {\em THERMAL SCIENCE}, 20(2):763--769, 2016.

\bibitem{atangana2018fractional}
Abdon Atangana and JF~G{\'o}mez-Aguilar.
\newblock Fractional derivatives with no-index law property: application to
  chaos and statistics.
\newblock {\em Chaos, Solitons \& Fractals}, 114:516--535, 2018.

\bibitem{beck2001dynamical}
Christian Beck.
\newblock Dynamical foundations of nonextensive statistical mechanics.
\newblock {\em Physical Review Letters}, 87(18):180601, 2001.

\bibitem{beck2009recent}
Christian Beck.
\newblock Recent developments in superstatistics.
\newblock {\em Brazilian Journal of Physics}, 39(2A):357--363, 2009.

\bibitem{beck2003superstatistics}
Christian Beck and Ezechiel~GD Cohen.
\newblock Superstatistics.
\newblock {\em Physica A: Statistical mechanics and its applications},
  322:267--275, 2003.

\bibitem{bressloff2014stochastic}
Paul~C Bressloff.
\newblock {\em Stochastic processes in cell biology}, volume~41.
\newblock Springer, 2014.

\bibitem{chechkin2017brownian}
Aleksei~V Chechkin, Flavio Seno, Ralf Metzler, and Igor~M Sokolov.
\newblock Brownian yet non-gaussian diffusion: from superstatistics to
  subordination of diffusing diffusivities.
\newblock {\em Physical Review X}, 7(2):021002, 2017.

\bibitem{cohen2004superstatistics}
EGD Cohen.
\newblock Superstatistics.
\newblock {\em Physica D: Nonlinear Phenomena}, 193(1-4):35--52, 2004.

\bibitem{dos2019analytic}
Maike A~F dos Santos.
\newblock Analytic approaches of the anomalous diffusion: A review.
\newblock {\em Chaos, Solitons \& Fractals}, 124:86--96, 2019.

\bibitem{dos2019fractional}
Maike A~F dos Santos.
\newblock Fractional prabhakar derivative in diffusion equation with non-static
  stochastic resetting.
\newblock {\em Physics}, 1(1):40--58, 2019.

\bibitem{dos2019continuous}
Maike A~F dos Santos.
\newblock From continuous-time random walks to controlled-diffusion reaction.
\newblock {\em Journal of Statistical Mechanics: Theory and Experiment},
  2019(3):033214, 2019.

\bibitem{dos2019mittag}
Maike A~F dos Santos.
\newblock Mittag--leffler memory kernel in l{\'e}vy flights.
\newblock {\em Mathematics}, 7(9):766, 2019.

\bibitem{fernandez2019series}
Arran Fernandez, Dumitru Baleanu, and HM~Srivastava.
\newblock Series representations for fractional-calculus operators involving
  generalised mittag-leffler functions.
\newblock {\em Communications in Nonlinear Science and Numerical Simulation},
  67:517--527, 2019.

\bibitem{garcia2011superstatistics}
Vladimir Garc{\'\i}a-Morales and Katharina Krischer.
\newblock Superstatistics in nanoscale electrochemical systems.
\newblock {\em Proceedings of the National Academy of Sciences},
  108(49):19535--19539, 2011.

\bibitem{garra2014hilfer}
Roberto Garra, Rudolf Gorenflo, Federico Polito, and {\v{Z}}ivorad Tomovski.
\newblock Hilfer--prabhakar derivatives and some applications.
\newblock {\em Applied mathematics and computation}, 242:576--589, 2014.

\bibitem{gomez2019time}
JF~G{\'o}mez-Aguilar and Abdon Atangana.
\newblock Time-fractional variable-order telegraph equation involving operators
  with mittag-leffler kernel.
\newblock {\em Journal of Electromagnetic Waves and Applications},
  33(2):165--177, 2019.

\bibitem{gorenflo2014mittag}
Rudolf Gorenflo, Anatoli~Aleksandrovich Kilbas, Francesco Mainardi, Sergei~V
  Rogosin, et~al.
\newblock {\em Mittag-Leffler functions, related topics and applications},
  volume~2.
\newblock Springer, 2014.

\bibitem{han2013gamma}
Jung~Hun Han.
\newblock Gamma function to beck--cohen superstatistics.
\newblock {\em Physica A: Statistical Mechanics and its Applications},
  392(19):4288--4298, 2013.

\bibitem{hanel2011generalized}
Rudolf Hanel, Stefan Thurner, and Murray Gell-Mann.
\newblock Generalized entropies and the transformation group of
  superstatistics.
\newblock {\em Proceedings of the National Academy of Sciences},
  108(16):6390--6394, 2011.

\bibitem{hapca2008anomalous}
Simona Hapca, John~W Crawford, and Iain~M Young.
\newblock Anomalous diffusion of heterogeneous populations characterized by
  normal diffusion at the individual level.
\newblock {\em Journal of the Royal Society Interface}, 6(30):111--122, 2008.

\bibitem{haubold2011mittag}
Hans~J Haubold, Arak~M Mathai, and Ram~K Saxena.
\newblock Mittag-leffler functions and their applications.
\newblock {\em Journal of Applied Mathematics}, 2011, 2011.

\bibitem{hristov2019linear}
Jordan Hristov.
\newblock Linear viscoelastic responses and constitutive equations in terms of
  fractional operators with non-singular kernels.
\newblock {\em The European Physical Journal Plus}, 134(6):283, 2019.

\bibitem{jaynes1957information}
Edwin~T Jaynes.
\newblock Information theory and statistical mechanics.
\newblock {\em Physical review}, 106(4):620, 1957.

\bibitem{kilbas2004generalized}
Anatoly~A Kilbas, Megumi Saigo, and Ram~K Saxena.
\newblock Generalized mittag-leffler function and generalized fractional
  calculus operators.
\newblock {\em Integral Transforms and Special Functions}, 15(1):31--49, 2004.

\bibitem{mathai2010mittag}
AM~Mathai and Hans~J Haubold.
\newblock Mittag-leffler functions to pathway model to tsallis statistics.
\newblock {\em Integral Transforms and Special Functions}, 21(11):867--875,
  2010.

\bibitem{mathai2012pathway}
AM~Mathai and Panagis Moschopoulos.
\newblock A pathway idea for model building.
\newblock {\em Journal of statistics applications \& probability}, 1(1):15,
  2012.

\bibitem{mathai2011pathway}
Arak~M Mathai and Hans~J Haubold.
\newblock A pathway from bayesian statistical analysis to superstatistics.
\newblock {\em Applied Mathematics and Computation}, 218(3):799--804, 2011.

\bibitem{metzler2016non}
R~Metzler, J-H Jeon, and AG~Cherstvy.
\newblock Non-brownian diffusion in lipid membranes: Experiments and
  simulations.
\newblock {\em Biochimica et Biophysica Acta (BBA)-Biomembranes},
  1858(10):2451--2467, 2016.

\bibitem{metzler2000random}
Ralf Metzler and Joseph Klafter.
\newblock The random walk's guide to anomalous diffusion: a fractional dynamics
  approach.
\newblock {\em Physics reports}, 339(1):1--77, 2000.

\bibitem{metzner2015superstatistical}
Claus Metzner, Christoph Mark, Julian Steinwachs, Lena Lautscham, Franz
  Stadler, and Ben Fabry.
\newblock Superstatistical analysis and modelling of heterogeneous random
  walks.
\newblock {\em Nature communications}, 6:7516, 2015.

\bibitem{oliveira2019anomalous}
Fernando~A Oliveira, Rogelma Ferreira, Luciano~C Lapas, and Mendeli~H
  Vainstein.
\newblock Anomalous diffusion: A basic mechanism for the evolution of
  inhomogeneous systems.
\newblock {\em Frontiers in Physics}, 7, 2019.

\bibitem{ourabah2018fractional}
Kamel Ourabah and Mouloud Tribeche.
\newblock Fractional superstatistics from a kinetic approach.
\newblock {\em Physical Review E}, 97(3):032126, 2018.

\bibitem{podlubny1998fractional}
Igor Podlubny.
\newblock {\em Fractional differential equations: an introduction to fractional
  derivatives, fractional differential equations, to methods of their solution
  and some of their applications}, volume 198.
\newblock Elsevier, 1998.

\bibitem{prabhakar1971singular}
Tilak~Raj Prabhakar et~al.
\newblock A singular integral equation with a generalized mittag leffler
  function in the kernel.
\newblock 1971.

\bibitem{prugel1994analysis}
Adam Pr{\"u}gel-Bennett and Jonathan~L Shapiro.
\newblock Analysis of genetic algorithms using statistical mechanics.
\newblock {\em Physical Review Letters}, 72(9):1305, 1994.

\bibitem{rouse2017superstatistical}
I~Rouse and S~Willitsch.
\newblock Superstatistical energy distributions of an ion in an ultracold
  buffer gas.
\newblock {\em Physical review letters}, 118(14):143401, 2017.

\bibitem{rudolf2000applications}
Hilfer Rudolf.
\newblock {\em Applications of fractional calculus in physics}.
\newblock world scientific, 2000.

\bibitem{saad2018analysis}
Khaled~M Saad and JF~G{\'o}mez-Aguilar.
\newblock Analysis of reaction--diffusion system via a new fractional
  derivative with non-singular kernel.
\newblock {\em Physica A: Statistical Mechanics and its Applications},
  509:703--716, 2018.

\bibitem{samko1993fractional}
Stefan~G Samko, Anatoly~A Kilbas, Oleg~I Marichev, et~al.
\newblock {\em Fractional integrals and derivatives}, volume 1993.
\newblock Gordon and Breach Science Publishers, Yverdon Yverdon-les-Bains,
  Switzerland, 1993.

\bibitem{sandev2018models}
Trifce Sandev, Weihua Deng, and Pengbo Xu.
\newblock Models for characterizing the transition among anomalous diffusions
  with different diffusion exponents.
\newblock {\em Journal of Physics A: Mathematical and Theoretical},
  51(40):405002, 2018.

\bibitem{sebastian2015overview}
Nicy Sebastian, Seema S~Nair, and Dhannya P~Joseph.
\newblock An overview of the pathway idea and its applications in statistical
  and physical sciences.
\newblock {\em axioms}, 4(4):530--553, 2015.

\bibitem{sene2019mittag}
Ndolane Sene.
\newblock Mittag-leffler input stability of fractional differential equations
  and its applications.
\newblock {\em Discrete \& Continuous Dynamical Systems-S}, pages 636--643,
  2019.

\bibitem{sene2019analysis}
Ndolane Sene and Karima Abdelmalek.
\newblock Analysis of the fractional diffusion equations described by
  atangana-baleanu-caputo fractional derivative.
\newblock {\em Chaos, Solitons \& Fractals}, 127:158--164, 2019.

\bibitem{sevilla2019stationary}
Francisco~J Sevilla, Alejandro~V Arzola, and Enrique~Puga Cital.
\newblock Stationary superstatistics distributions of trapped run-and-tumble
  particles.
\newblock {\em Physical Review E}, 99(1):012145, 2019.

\bibitem{sposini2018random}
Vittoria Sposini, Aleksei~V Chechkin, Flavio Seno, Gianni Pagnini, and Ralf
  Metzler.
\newblock Random diffusivity from stochastic equations: comparison of two
  models for brownian yet non-gaussian diffusion.
\newblock {\em New Journal of Physics}, 20(4):043044, 2018.

\bibitem{tolman1979principles}
Richard~Chace Tolman.
\newblock {\em The principles of statistical mechanics}.
\newblock Courier Corporation, 1979.

\bibitem{touchette2005asymptotics}
Hugo Touchette and Christian Beck.
\newblock Asymptotics of superstatistics.
\newblock {\em Physical Review E}, 71(1):016131, 2005.

\bibitem{tsallis1988possible}
Constantino Tsallis.
\newblock Possible generalization of boltzmann-gibbs statistics.
\newblock {\em Journal of statistical physics}, 52(1-2):479--487, 1988.

\bibitem{tsallis2003constructing}
Constantino Tsallis and Andre~MC Souza.
\newblock Constructing a statistical mechanics for beck-cohen superstatistics.
\newblock {\em Physical Review E}, 67(2):026106, 2003.

\bibitem{YALCIN20135431}
G~Cigdem Yalcin and Christian Beck.
\newblock Environmental superstatistics.
\newblock {\em Physica A: Statistical Mechanics and its Applications},
  392(21):5431--5452, 2013.

\end{thebibliography}






% \end{thebibliography}

\end{document}